\documentclass[secnumarabic,amssymb, nobibnotes, aps, prd, superscriptaddress,showpacs]{revtex4}
\usepackage[dvips]{graphicx}

\begin{document}

\title{{\it Ab initio} calculation of H + He$^+$ charge transfer cross sections for plasma physics}%

\author{J. Loreau}
\affiliation{Laboratoire de Chimie Quantique et Photophysique, Universit\'e Libre de Bruxelles, CP160/09 50, av. F. D. Roosevelt, 1050 Bruxelles, Belgium}
\author{K. Sodoga}
\affiliation{Laboratoire de Chimie Physique, B\^at 349, Universit\'e de Paris-Sud, UMR8000, Orsay, F-91405, France.}
\affiliation{Facult\'e des Sciences, D\'epartement de Physique, Universit\'e de Lom\'e, BP 1515 Lom\'e, Togo}
\author{D. Lauvergnat}
\affiliation{Laboratoire de Chimie Physique, B\^at 349, Universit\'e de Paris-Sud, UMR8000, Orsay, F-91405, France.}
\author{M. Desouter-Lecomte}
\affiliation{Laboratoire de Chimie Physique, B\^at 349, Universit\'e de Paris-Sud, UMR8000, Orsay, F-91405, France.}
\affiliation{D\'epartement de Chimie, B6c, Universit\'e de Li\`ege, Sart-Tilman, B-4000, Li\`ege 1, Belgium}
\author{N. Vaeck}
\affiliation{Laboratoire de Chimie Quantique et Photophysique, Universit\'e Libre de Bruxelles, CP160/09 50, av. F. D. Roosevelt, 1050 Bruxelles, Belgium}


\begin{abstract}
The charge transfer in low energy (0.25 to 150 eV/amu) H($nl$) + He$^+(1s)$ collisions is investigated using a quasi-molecular approach for the $n=2,3$ as well as the first two $n=4$ singlet states.
The diabatic potential energy curves of the HeH$^+$ molecular ion are obtained from the adiabatic potential energy curves and the non-adiabatic radial coupling matrix elements using a two-by-two diabatization method, and a time-dependent wave-packet approach is used to calculate the state-to-state cross sections. 
We find a strong dependence of the charge transfer cross section in the principal and orbital quantum numbers $n$ and $l$ of the initial or final state. We estimate the effect of the non-adiabatic rotational couplings, which is found to be important even at energies below 1 eV/amu. However, the effect is small on the total cross sections at energies below 10 eV/amu. We observe that to calculate charge transfer cross sections in a $n$ manifold, it is only necessary to include states with $n^{\prime}\leq n$, and we discuss the limitations of our approach as the number of states increases.

\end{abstract}


\maketitle

\section{Introduction}

Starting with the historical work of Massey and Smith in 1933 \cite{Massey1933} and due to the apparent simplicity of this two electron system, the asymmetrical charge transfer process He ($1s^{2}\ ^1S$) + H$^{+}$  $\longrightarrow$ H + He$^{+}$ has quickly been considered as a prototype for semi-classical methods to treat collisions \cite{Green1965, Bransden1966,SinFaiLam1967, Colegrave1968, Winter1976}. This process is dominated by the capture into the H($1s$) state but later on, the charge transfer excitation and the direct excitation processes have been studied in detail both theoretically and experimentally \cite{Kimura1985, Kimura1986, Jain1987,Fritsch1988,Shah1989,Slim1990,Slim1991,Jackson1992,Shingal1991}. All these reactions require intermediate collision energy which suits perfectly the semi-empirical description of the collision. It is also the case for the charge transfer mechanism that involves H$^{-}$ ion as projectile (He$^{2+}$ + H$^{-}$  $\rightarrow$ H + He$^{+}(nl)$) \cite{Terao1988,Cherkani1991}. 

At much lower collisional energy, charge transfer can populate excited states of He($1snl\ ^{1,3}L$) from the corresponding excited states $nl$ of H. Although there is no measurement for those processes, they have been studied theoretically using a semi-classical approach with a linear trajectory for the nuclei taking into account the coupling between the final Stark splitting H states and the initial He states at large internuclear distances ($R\geq 20$ a.u.) and neglecting electron translational factors and rotational couplings \cite{Chibisov2001,Chibisov2002}. These works provide data for an energy range between 2.5 eV/amu and 10 keV/amu. 

Recently, it has appeared that those low energy charge transfer processes involving excited hydrogen states could be of major importance for the monitoring of warm plasmas \cite{Rosmej2006}. Indeed, spectroscopic methods are among the most effective approaches to determine particle transport in magnetic confined fusion plasmas. From this point of view, simulations of excited He emissions resolved in space and time has been proposed as a tool independent of the theoretical plasma model.  Atomic physics simulations can be compared to experimental data and the diffusion and convective velocity parameters could be determined. However, the He ions will interact with the H/D background via a charge transfer mechanism which modifies the population of the He excited states and therefore the intensity of the emission lines. A self-consistent approach to the description of the coupling of the radiating He with the plasma background via charge transfer has shown that these processes are very important at low collisional energies (typically of the order of 0.1 eV to 100 eV) and that, therefore, an accurate knowledge of charge transfer cross sections in this energy range is essential \cite{Rosmej2006}.

In this work, we have adopted a quasi-molecular approach of the ion-atom collision based on the use of quantum-chemistry {\it ab initio} methods to obtain the potential energy curves as well as the radial and rotational coupling matrix elements of the quasi-molecule HeH$^{+}$. A wave packet method is used to treat the curve-crossing dynamics resulting from the failure of the Born-Oppenheimer approximation. A Gaussian wave packet is prepared in the entrance channel and propagated on the coupled ro-electronic channels. The collision matrix elements are computed from an analysis of the flux in the asymptotic region by using properties of absorbing potentials, giving access to the charge transfer cross-sections for the processes He$^{+}$ ($1s$) + H($nl$) $\rightarrow$ He ($1sn^{\prime}l^{\prime}\  ^{1,3}L$) + H$^{+}$ where $n=2-3$. We estimate the influence of the rotational couplings on the cross section.

\section{Theory}

\subsection{Molecular data}

The Hamiltonian is given as the sum of an electronic part and a nuclear kinetic part:
\begin{equation}
H=T^{\mathrm{N}}+H^{\mathrm{el}}
\end{equation}

The electronic Hamiltonian includes a kinetic term for the electrons and all the potential energy terms.

The potential energy curves (PEC) $U_{m\Lambda}$ and the adiabatic electronic functions $\zeta_{m\Lambda}$ solve the electronic motion:
\begin{equation}
H^{\mathrm{el}}\zeta_{m\Lambda}({\bf r}; R)=U_{m\Lambda}(R)\zeta_{m\Lambda}({\bf r}; R) \ ,
\end{equation} 
where ${\bf r}$ stands for the electron coordinates and $R$  is the radial coordinate for the nuclei. $m$ is used to number the states for a given $\Lambda$, which is the quantum number associated to $L_z$, the projection of the total electronic orbital angular momentum ${\bf L}$ onto the molecular $z$ axis. The molecular electronic states are classified according to the value of $\vert\Lambda\vert$: $\Sigma$ states correspond to $\Lambda=0$, $\Pi$ states to $\vert\Lambda\vert=1$, and $\Delta$ states to $\vert\Lambda\vert=2$. We therefore see that states with $\vert \Lambda\vert \neq 0$ are doubly degenerate for singlet states. In the atomic limit ($R\rightarrow\infty$), $\Lambda$ becomes $m_L$, the magnetic quantum number.

On the other hand, $T^{\mathrm{N}}$ can be written in atomic units as the sum of a radial part,
\begin{equation}
H^{\mathrm{rad}}=-\frac{1}{2\mu}\partial_{R}^2 \ ,
\end{equation}
where $\mu$ is the reduced mass of the system, and a rotational part given by
\begin{equation}
H^{\mathrm{rot}} = \frac{1}{2\mu R^2}{\bf N}^2 = \frac{1}{2\mu R^2}\Big[{\bf K}^2 + {\bf L}^2 -2K_{z}L_{z} - K_{+}L_{-} - K_{-}L_{+}\Big]
\end{equation}
where ${\bf N}$ is the nuclear angular momentum. In this work, we will focus on singlet states and we will not consider any spin-dependent interactions, so that the total angular momentum is ${\bf K}={\bf N}+{\bf L}$.
Since $T^{\mathrm{N}}=H^{\mathrm{rad}}+H^{\mathrm{rot}}$, the nuclear wave function is the product of a radial part and an angular part: $\psi_{m\Lambda}({\bf R}) = \psi_{m\Lambda}(R)\ \vert K\Lambda M\rangle$. The angular functions are eigenfunctions of ${\bf K}^2$ and $K_{z}$ with eigenvalues $K(K+1)$ and $\Lambda$, respectively. The action of the ladder operators $K_{\pm}=K_x \pm iK_y$ is given by $K_{\pm}\vert K\Lambda  M\rangle= \big[ K(K+1)-\Lambda(\Lambda\mp 1)\big]^{1/2}\vert K\Lambda\mp 1 M\rangle$. 

In the basis of these electronic-rotational functions, the matrix elements of $H^{\mathrm{rot}}$ are given by
\begin{eqnarray}\label{hrot}
H^{\mathrm{rot}}_{m\Lambda K, m^{\prime}\Lambda^{\prime}K^{\prime}}  =   \frac{1}{2\mu R^2}  \bigg\{ \Big( K(K+1) & - \Lambda^2 \Big)\delta_{mm^{\prime}}\delta_{\Lambda\Lambda^{\prime}}  - (L_{-})_{mm^{\prime}} \left[ K(K+1)-\Lambda(\Lambda- 1)\right]^{1/2} \delta_{\Lambda,\Lambda^{\prime}-1} \nonumber\\
& - (L_{+})_{mm^{\prime}}\big[ K(K+1)-\Lambda(\Lambda + 1) \big]^{1/2}  \delta_{\Lambda,\Lambda^{\prime}+1} \bigg\} \delta_{KK^{\prime}} 
\end{eqnarray}
where  the contribution from $(L_{x}^2+L_{y}^2)_{mm^{\prime}} \delta_{\Lambda\Lambda^{\prime}}$ has been neglected. We see that states with $\Delta\Lambda=\pm1$ will interact through the rotational Hamiltonian.

To treat the effects of the rotational Hamiltonian, it is more convenient to work with parity adapted functions \cite{Lefebvre-Brion2004}. These functions are defined by
\begin{equation}\label{parity-adapted}
\vert m K \Lambda  M\epsilon \rangle = \frac{1}{\sqrt{2+2\delta_{\Lambda 0}}} \Big[ \vert K\Lambda M\rangle \zeta_{m\Lambda} + (-1)^K \epsilon \vert K -\Lambda M\rangle \zeta_{m-\Lambda} \Big]
\end{equation}
 where $\epsilon =1$ and $\epsilon=-1$ correspond to $e$ and $f$ states, respectively. 
Using (\ref{hrot}) and (\ref{parity-adapted}), it can be shown that $H^{\mathrm{rot}}$ only connects states of the same parity. As $^1\Sigma^+$ states can have $e$ or $f$ symmetry, 	only half the $\Pi$ states must be taken into account in the calculations.



\begin{figure}[h!]
\centering
\hspace{-1cm}
\includegraphics[angle=-90,width=15cm]{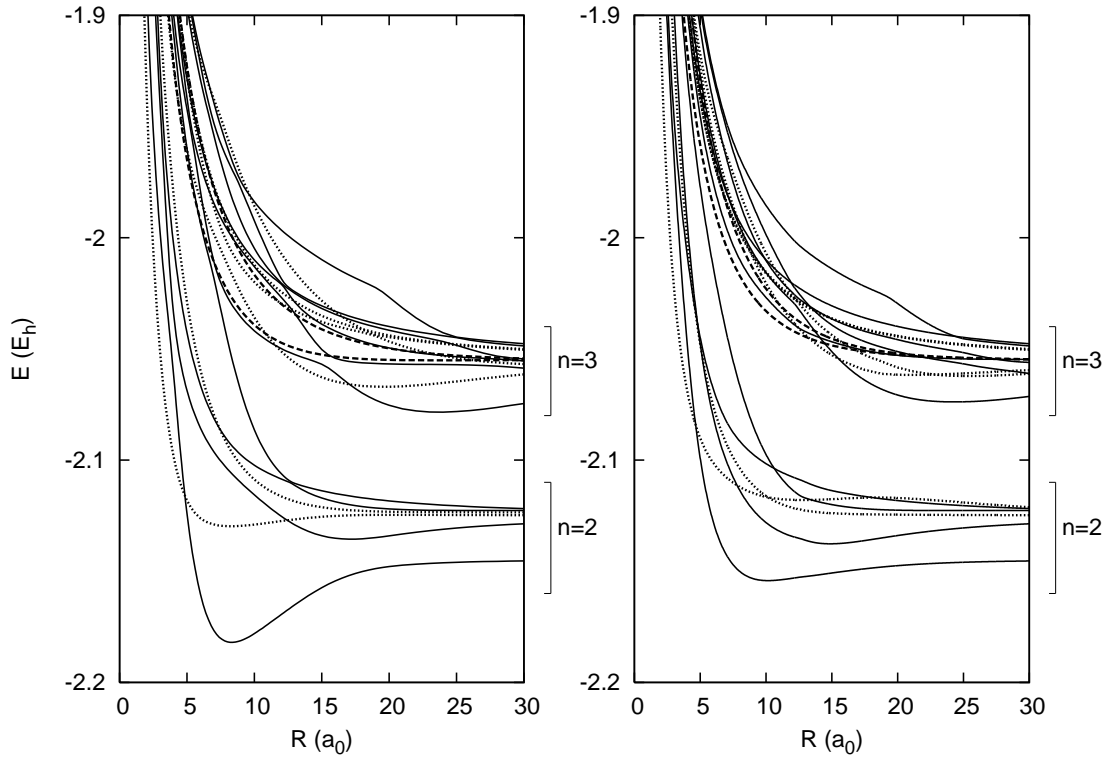}
\caption{Adiabatic (left) and diabatic (right) PEC of the $n=2,3$ states of HeH$^+$. Full lines, $^1\Sigma^+$ states. Dots, $^1\Pi$ states. Dashed lines, $^1\Delta$ states. The $n=1$ $^1\Sigma^+$ states have been excluded from the diabatization procedure.}
\label{PEC}
\end{figure}

We will consider here the $n=1-3$ $^1\Sigma^+$, $^1\Pi$ and $^1\Delta$ states as well as the first two $n=4$ $^1\Sigma^+$ states. We could not include more $n=4$ states in the calculations, as we were not able to calculate the radial non-adiabatic couplings between these states by {\it ab initio} methods. The dissociative atomic states and the asymptotic energies of the molecular states are shown in table \ref{table_states}. We have also considered the $n=2$ $^3\Sigma^+$ and $^3\Pi$ states to allow the comparison with \cite{Chibisov2001,Chibisov2002} (see section \ref{CS_tot}). The adiabatic potential energy curves (PEC) for these states have been calculated using the {\it ab initio} quantum chemistry package MOLPRO version 2006.1 \cite{Molpro}. An adapted basis set consisting of the aug-cc-pv5Z basis set \cite{Dunning1989} supplemented by one contracted Gaussian function per orbital per atom up to $n=4$ has been used. Details of the calculations can be found in \cite{Loreau2010}. Due to the electric field produced by He$^+(1s)$, there is a Stark mixing of the hydrogen states. However, these Stark states adiabatically become pure atomic states as $R\rightarrow\infty$ so that we know with certainty the atomic configuration corresponding to a molecular state, as indicated in table \ref{table_states}.
The PEC have been calculated at the state-averaged complete active space self-consistent field (CASSCF) level and are shown in figure \ref{PEC} for the $n=2-3$ states. This approach allows us to compute the radial non-adiabatic coupling matrix elements $F_{m\Lambda,m^{\prime}\Lambda}=\langle\zeta_{m\Lambda} \vert \partial_{R} \vert \zeta_{m^{\prime}\Lambda} \rangle$, which are used to build the diabatic representation \cite{Smith1969}. The adiabatic-to-diabatic transformation matrix $\mathbb{D}$ is the solution to the differential matrix equation $\partial_R\mathbb{D}+\mathbb{F\cdot D}=0$ and the diabatic potential energy curves are the diagonal elements of the matrix $\mathbb{U}^\mathrm{d}=\mathbb{D}^{-1}\cdot \mathbb{U\cdot D}$, where $\mathbb{U}$ is the matrix of $H^{\mathrm{el}}$ in the adiabatic representation.
We have used an approximate $\mathbb{F}$ matrix by keeping only the couplings between adjacent states, {\it i.e.} the elements $F_{m,m+1}$. The main reason is that this approach simplifies considerably the diabatization procedure \cite{Loreau2010}. It has been shown to give similar results as when the complete $\mathbb{F}$ matrix is taken into account in dynamical calculations \cite{Zhu1997}, something we have also observed in the low-energy calculation of the charge transfer cross sections for the $n=2$ $^1\Sigma^+$ states (see below).

\begin{table}
\caption{\label{table_states} CASSCF energies at $R=70$ a.u. and dissociative products of the singlet states included in the calculations.}
\begin{tabular}{ccccl}
$^{2S+1}\Lambda$	& $n$	& $m$ & Energy (hartree)	& Dissociative atomic states  \\
			&	& & at $R=70$ a.u.	&  \\ \hline
$^1\Sigma^+$	&	$n=1$ 	& 1	& -2.90324307 & He($1s^2\ ^1S$) + H$^+$ \\
			&			& 2	& -2.49995502 & He$^+(1s)$ + H($1s$) \\ 
			&	 $n=2$	& 3	& -2.14589424 & He($1s2s\ ^1S$) + H$^+$ \\
			&	 		& 4	& -2.12556499 & He$^+(1s)$ + H($2p$)  \\
			&	 		& 5	& -2.12433765 & He$^+(1s)$ + H($2s$)  \\
			&	 		& 6	& -2.12374055 & He($1s2p\ ^1P^{o}$) + H$^+$ \\
			&	 $n=3$	& 7	& -2.06157066 & He($1s3s\ ^1S$) + H$^+$ \\
			&	 		& 8	& -2.05758300 & He$^+(1s)$ +  H($3d$) \\ 
			&	 		& 9	& -2.05632793 & He($1s3d\ ^1D$) + H$^+$ \\
			&	 		& 10 & -2.05537040 & He$^+(1s)$ + H($3p$) \\
			&	 		& 11 & -2.05379172 & He($1s3p\ ^1P^{o}$) + H$^+$ \\
			&	 		& 12 & -2.05411889 & He$^+(1s)$ + H($3s$)  \\
			&	$n=4$	& 13	& -2.03701879 & He($1s4s\ ^1S$) + H$^+$ \\
			&			& 14 & -2.03502013 & He$^+(1s)$ + H($4p$)  \vspace{0.2cm} \\ 
$^1\Pi$		&	$n=2$	& 1	& -2.12491660 & He$^+(1s)$ + H($2p$)  \\
			&	 		& 2	& -2.12368473 & He($1s2p\ ^1P^{o}$) + H$^+$ \\ 
			&	$n=3$	& 3	& -2.05639837 & He$^+(1s)$ + H($3d$) \\
			&	 		& 4	& -2.05616425 & He($1s3d\ ^1D$) + H$^+$ \\
			&	 		& 5	& -2.05456333 & He$^+(1s)$ + H($3p$) \\
			&	 		& 6	& -2.05419837 & He($1s3p\ ^1P^{o}$) + H$^+$ \vspace{0.2cm} \\
$^1\Delta$	&	$n=3$	& 1	& -2.05540649 & He$^+(1s)$ + H($3d$)  \\
			&			& 2	& -2.05537712 & He($1s3d\ ^1D$) + H$^+$ \vspace{0.2cm} \\
\hline
\end{tabular}
\end{table}


The non-adiabatic rotational coupling matrix elements $\langle \zeta_{m\Lambda} \vert L_\pm \vert \zeta_{m^{\prime}\Lambda^{\prime}}\rangle$ appearing in equation (\ref{hrot}) have also been computed at the CASSCF level using MOLPRO. As pointed out in \cite{Belyaev2001, Belyaev2002} and \cite{Loreau2010}, some of these couplings behave asymptotically as $R$, which is due to the choice of the electronic coordinates. Due to the factor $1/R^2$ in equation (\ref{hrot}), these couplings will thus decrease as $1/R$, much slower that the radial couplings which decrease to zero extremely fast outside the interaction region. This causes a problem in the calculation of the cross sections as it implies the use of very large numerical grids that increase tremendously the calculation time. To solve this problem, we modified the problematic rotational couplings outside the interaction region where we required that they decrease to the atomic values of the couplings. We have tried various switching functions to find a set of parameters that had no effect on the cross sections. This approximation is also justified in our case by the fact that the linear rotational couplings usually connect two states in the same atomic configuration, so that the modification will not influence the charge transfer cross sections.

There are also cases where the atomic value of $\langle \zeta_{m\Lambda} \vert L_\pm \vert \zeta_{m^{\prime}\Lambda^{\prime}}\rangle$ is a constant, but not zero. The rotational Hamiltonian then decreases as $1/R^2$, which still implies the use of large numerical grids. However, this can only happen again for transitions between two states in the same atomic configuration (electron excitation), a process we do not consider here.

In the atomic limit, when the non-adiabatic rotational couplings are neglected, our method implies conservation of the magnetic quantum number $m_L$ as the Hamiltonian is diagonal in $\Lambda$. When they are included in the calculations, we have interaction between states with $\Delta m_L = 0,\pm 1$.

\subsection{Cross section calculation}

The cross section corresponding to the transfer of an electron from an initial state $m,\Lambda$ to a final state $m^{\prime},\Lambda^{\prime}$ is given by \cite{Child1974}
\begin{equation}\label{QCSsum}
\sigma_{m^{\prime}\Lambda^{\prime},m\Lambda}(E) = \frac{\pi}{k^2_{m\Lambda}(E)}\sum_{K}(2K+1)\vert S^K_{m^{\prime}\Lambda^{\prime},m\Lambda}(E)-\delta_{m^{\prime}m}\delta_{\Lambda^{\prime}\Lambda}\vert^2
\end{equation}
where $k_{m\Lambda}$ is the wave number in the entrance channel, $k_{m\Lambda}=\sqrt{2\mu(E-U_{m\Lambda })}$. 
As the Hamiltonian is diagonal in $K$ (see equation (\ref{hrot})), the cross section must be calculated for each value of $K$ until convergence in equation (\ref{QCSsum}).

We use the coupled channel formalism in the rotational-electronic diabatic representation. In the time-dependent formalism, we start by defining a Gaussian initial wave packet which is propagated in time using the split operator algorithm \cite{Fleck1976}. The coupled equations give access to the wave packets on all the rotational-electronic states. For each value of $K$, the scattering matrix elements $\vert S^K_{m^{\prime}\Lambda^{\prime},m\Lambda}(E)\vert^2$ are then extracted using the flux operator formalism with a complex absorbing potential \cite{Jackle1996,Baloitcha2001}.

We start by defining the functions
\begin{equation}\label{hankel}
\Phi^{\pm,K}_{m\Lambda,E}=\sqrt{\frac{\mu}{2\pi k_{m\Lambda}}}h^{\pm}_{K}(k_{m\Lambda}R)\zeta_{m\Lambda}^{\mathrm{d}}
\end{equation}
where $h^{\pm}_{K}(k_{m\Lambda}R)$ are the Riccati--Hankel functions \cite{Morrison2007} and $\zeta_{m\Lambda}^{\mathrm{d}}$ are the electronic wave functions in the diabatic representation.

We then introduce the time-independent energy normalized wave functions $\vert\Psi^{+,K}_{m\Lambda,E}\rangle$, solutions of
\begin{equation}
\Big(-\frac{1}{2\mu}\partial_{R}^2  + \frac{1}{2\mu R^2} (K(K+1) - \Lambda^2) + H^{\mathrm{el}} \Big) \vert\Psi^{+,K}_{m\Lambda,E}\rangle = U_{m\Lambda}\vert\Psi^{+,K}_{m\Lambda,E}\rangle
\end{equation}
and satisfying the asymptotic condition
\begin{equation}\label{boundary-cond}
\vert\Psi^{+,K}_{m\Lambda,E}\rangle \quad \stackrel{R\rightarrow \infty}{\longrightarrow} \quad \vert \Phi^{-,K}_{m\Lambda,E} \rangle - \sum_{m^{\prime},\Lambda^{\prime}} S_{m^{\prime}\Lambda^{\prime},m \Lambda}^K(E) \vert\Phi^{+,K}_{m^{\prime}\Lambda ^{\prime},E}\rangle
\end{equation}
These stationary eigenfunctions can be constructed as the Fourier-transform of a time-dependent wave packet $\Phi (t)$:
\begin{equation}\label{Psi-FT}
\vert\Psi^{+,K}_{m\Lambda,E}\rangle = \frac{1}{2\pi \Gamma_{m\Lambda}^K(E)}\int_{-\infty}^{+\infty} \vert\Phi_{m\Lambda} (t)\rangle \exp(iEt)\  dt\ .
\end{equation}
The vector $\vert\mathbf{\Phi}(t)\rangle$ is constructed by propagating an initial wave packet $\vert \mathbf{\Phi}(0)\rangle$ in time using the Hamiltonian matrix in the rotational-electronic diabatic representation:
\begin{equation}\label{time-evolution}
\vert \mathbf{\Phi}(t)\rangle = \exp(-i\mathbb{H}^{\mathrm{d}}t)\vert \mathbf{\Phi}(0) \rangle \ .
\end{equation}
The initial wave packet is zero except in the diabatic channel $m\Lambda$, where it is represented by a Gaussian function $g(R)$ of width $\sigma$ and centered around $R_0$:
\begin{equation}
g(R) = \frac{1}{\sqrt{\sigma \sqrt{2/\pi}}} \exp\bigg[ik_{0}R - \frac{(R-R_{0})^2}{\sigma^2}\bigg]
\end{equation}
$\Gamma_{\Lambda m}^K$ is  the amplitude of the initial wave packet on the stationary states:
\begin{equation}
\Gamma_{m\Lambda}^K =  \langle \Psi^{+,K}_{m\Lambda,E}\vert \Phi_{0}\rangle
 = \sqrt{\frac{\mu}{2\pi k_{m\Lambda}}} \int_{0}^{\infty} h^{+}_{K}(k_{m\Lambda}R) g(R) dR \ .
\end{equation}

The flux operator is defined by \cite{Jackle1996}
\begin{equation}\label{flux}
F=-\frac{i}{2\mu}\Big(\frac{\partial}{\partial R}\delta (R-R_{c}) + \delta (R-R_{c})\frac{\partial}{\partial R}\Big)\ ,
\end{equation}
where $R_{c}$ is a point in the asymptotic region ({\it i.e.} such that there is no interactions for $R\geq R_{c}$) located behind $R_0$.

Using equations (\ref{hankel}), (\ref{boundary-cond}) and (\ref{flux}), one arrives at
\begin{equation}
\langle \Psi_{m\Lambda,E}^{+,K}\vert F \vert \Psi_{m\Lambda,E}^{+,K}\rangle = \frac{1}{2\pi}\sum_{m^{\prime},\Lambda^{\prime}}\vert S_{m^{\prime}\Lambda^{\prime},m\Lambda}^K(E)\vert ^{2}
\end{equation}
The sum can be removed using the projector onto the electronic state $m^{\prime}\Lambda^{\prime}$, $P_{m^{\prime}\Lambda^{\prime}}=\vert\zeta_{m^{\prime}\Lambda^{\prime}}^{\mathrm{d}}\rangle\langle\zeta_{m^{\prime}\Lambda^{\prime}}^{\mathrm{d}}\vert$, to obtain
\begin{equation} \label{flux2}
\langle \Psi_{m\Lambda,E}^{+,K}\vert P_{m^{\prime}\Lambda^{\prime}}FP_{m^{\prime}\Lambda^{\prime}} \vert \Psi_{m\Lambda,E}^{+,K}\rangle = \frac{1}{2\pi}\vert S_{m^{\prime}\Lambda^{\prime},m\Lambda}^K(E)\vert ^{2}
\end{equation}

A complex absorbing potential $-iW$ is then added to the Hamiltonian which becomes $H^{\prime}=H-iW$. 
The left hand side of equation (\ref{flux2}) is then calculated using equations (\ref{Psi-FT}) and (\ref{time-evolution}) with $H^{\prime}$ instead of $H$. This is allowed if the CAP is ``switched on'' in the asymptotic region: in the interaction region, the CAP vanishes and the value of $\Phi(t)$ propagated with $H$ or $H^{\prime}$ will be identical. Combining equations (\ref{Psi-FT}) and (\ref{flux2}), we find that the state-to-state cross section is given by

\begin{equation}\label{Smatrix}
\vert S_{m^{\prime}\Lambda^{\prime},m\Lambda}^K(E)\vert ^{2}=\frac{1}{2\pi \vert \Gamma_{m\Lambda}^K(E)\vert^2} \int_0^{\infty} dt \int_0^{\infty} dt^{\prime} \langle \Phi(t) \vert P_{m^{\prime}\Lambda^{\prime}} W P_{m^{\prime}\Lambda^{\prime}} \vert \Phi(t^{\prime}) \rangle \exp[iE(t^{\prime}-t)]
\end{equation}
Equation (\ref{Smatrix}) is used to obtain the matrix elements of $S$.

In our calculations, we used a CAP given by
\begin{equation}
W(R)=\eta_{c}\frac{(R-R_{c})^2}{R_{\infty}-R_{c}}
\end{equation}
where $\eta_{c}$ is the strength of the CAP and $R_{\infty}$ is the last point of the grid.


\section{Charge transfer cross sections}

\subsection{Computational details}

\subsubsection{Parameters for dynamics}

The parameters are chosen so as to ensure convergence of the sum in equation (\ref{QCSsum}) while keeping the norm of the $S$ matrix close to unity. For the calculations involving the $n=2$ states, a typical set of parameters consists of $2^{12}$ points for a grid of 60 a.u., an initial wave packet located around $R_{0}=40$ a.u. of width $\sigma=0.2$ and a CAP starting at $R_{c}=45$ a.u. of strength $\eta_{c}=0.01$. The time needed for the wave packet to return to the asymptotic region obviously depends on the collision energy. For the set of parameters above, it is approximatively contained between $2\cdot 10^3$ and $3\cdot 10^4$ a.u. for energies between 100 and $0.2$ eV/amu.

For the calculations of the cross sections involving $n=3$ states, we had to use grids up to $R=100$ a.u. This is due to the fact that the number of avoided crossings increases strongly with $n$, so that the positions of the radial non-adiabatic couplings are shifted to larger internuclear distances \cite{Loreau2010}. The time needed for the wave packet to return to the asymptotic region is therefore increased, and can be as high as $5\cdot 10^4$ a.u. for low energies.

When the rotational couplings are included in the calculations, the convergence of the partial cross sections is considerably slower as a function of $K$. It is therefore necessary to use much larger grids and the time of propagation is therefore increased tremendously. In addition, the number of points of the grid must also be increased to keep a constant step $dR$, again extending the computational time. In order to reduce the calculation time, we used grids of variable size ranging from 150 a.u. for small $K$ to 600 a.u. for values of $K$ around 2500. When treating the $n=3$ states, we could not perform calculations of partial cross sections for energies higher than 10 eV/amu as even these large grids did not ensure convergence.
The width of the wave packet was also increased up to $\sigma=1.5$ for these large grids, as wave packets with larger width stay more compact.

\subsubsection{Non-adiabatic radial couplings}

\begin{figure}[h!]
\centering
\includegraphics[width=12cm]{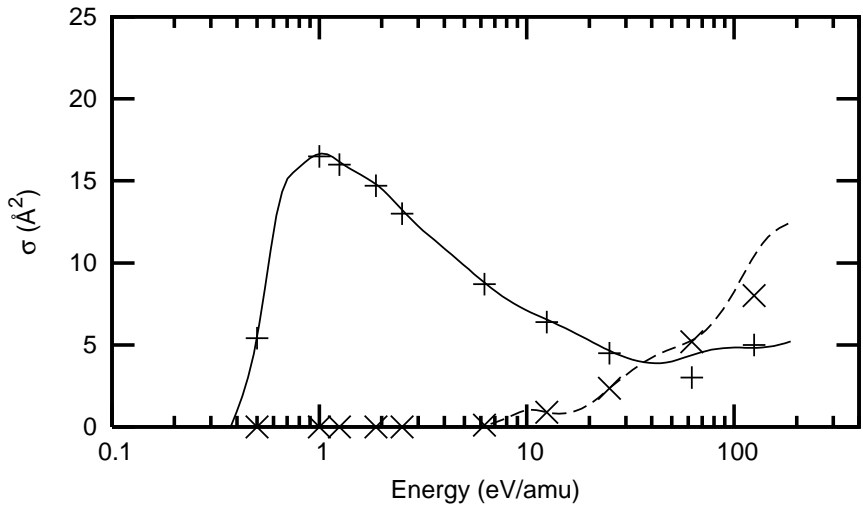}
\caption{Comparison of the two-by-two diabatization with the use of the complete $\mathbb{F}$ matrix in the calculation of charge transfer cross sections with He$^+(1s)$ + H$(2s)$ in the $^1\Sigma^+$ symmetry as the initial state. Full line and $+$ signs: charge transfer onto He$(1s2p\ ^1P^o)$ + H$^+$ with two-by-two or complete $\mathbb{F}$ matrix, respectively. Dashed line and $\times$ signs: charge transfer onto He$(1s2s\ ^1S)$ + H$^+$ with two-by-two or complete $\mathbb{F}$ matrix, respectively.}
\label{Comp_F}
\end{figure}

The first result that needs to be established is the validity of our approximation which consists in only retaining the non-adiabatic radial couplings $F_{m,m+1}$ instead of the complete $\mathbb{F}$ matrix. We show in figure \ref{Comp_F} a comparison between the two methods in the calculation of the cross sections for the process H$(2s)$ + He$^+(1s) \rightarrow$ H$^+$ + He$(1s2l\ ^1L)$. We conclude that this approximation is perfectly valid at low energies but that small deviations are observed at higher energies, $E\sim100$ eV/amu. The same conclusion is reached for the cross section for the process H$(2p)$ + He$^+(1s) \rightarrow$ H$^+$ + He$(1s2l\ ^1L)$.

Another issue is the fact that due to the Stark effect on hydrogen, some PEC undergo avoided crossings at large internuclear distances. However, as was pointed out in \cite{Loreau2010}, the large amplitude and the narrowness of the non-adiabatic radial couplings at those points indicate that a full diagonal diabatic representation at the crossing is perfectly justified, so that these crossings will not affect the cross sections.

Finally, as pointed out in \cite{Loreau2010}, the effect of the electron translation factors was found to be negligible in the range of energy considered in this work.

\subsection{General observations}


From a practical viewpoint, the calculation time of a cross section goes roughly as $e^{0.3m}$, where $m$ is the number of states included in the calculations, so that the computing time doubles every time two additional states are considered. This again extends the computational time when rotational interactions are taken into account, as $\Sigma$ and $\Pi$ states must be considered in the same calculation. It is therefore important to take the least states possible, and we will see that states with different values of $n$ can be considered independently.

\begin{figure}[h!]
\centering
\includegraphics[width=12cm]{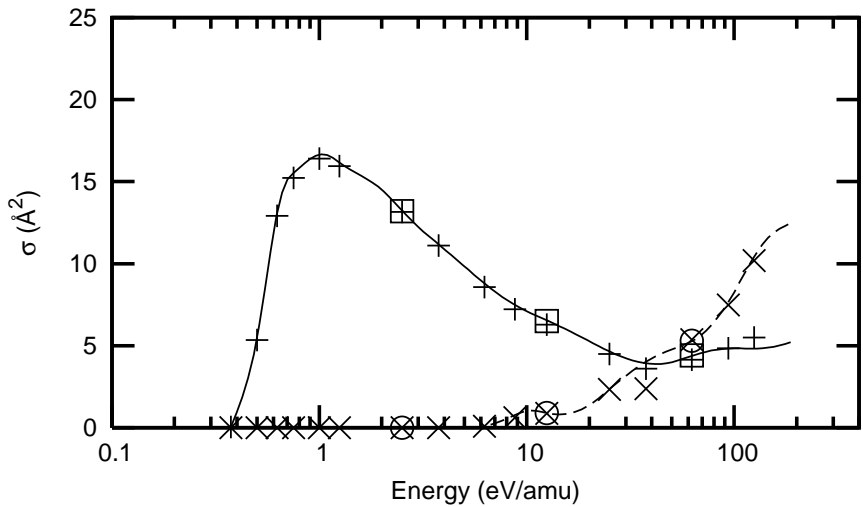}
\caption{Illustration of the possibility to treat states in different $n$ manifolds independently in the calculation of charge transfer cross sections with He$^+(1s)$ + H$(2s)$ in the $^1\Sigma^+$ symmetry as the initial state. 
Full line: charge transfer onto He$(1s2p\ ^1P^o)$ + H$^+$ using the four $n=2$ states. $+$ signs: the same, but with the two $n=1$ states included. $\square$ signs: the same, but with the six $n=3$ states included.
Dashed line: charge transfer onto He$(1s2s\ ^1S)$ + H$^+$ using the four $n=2$ states. $\times$ signs: the same, but with the two $n=1$ states included. $\bigcirc$ signs: the same, but with the six $n=3$ states included. }
\label{Comp_n}
\end{figure}

A few things seem to come out from our cross section calculations, which are presented below. The first is that, to a good degree of precision, the charge transfer cross sections in a given $n$ manifold are not modified by the inclusion of states with a principal quantum number $n^{\prime}\neq n$, a fact illustrated in figure \ref{Comp_n} for the $n=2$ $^1\Sigma^+$ states. It was anticipated that the $n=1$ states did not play any role in the $n>1$ cross sections since they are much lower in energy, but the possibility to treat the $n=2$ and $n=3$ states separately was much less obvious. Indeed, the shape of the diabatic PEC is strongly influenced by the inclusion of states with different principal quantum number and the density of states increases with $n$ \cite{Loreau2010}. We reached the same conclusion for the influence of the first two $n=4$ states on the $n=3$ manifold and in the $^1\Pi$ symmetry. 

However, while the cross section from the $n=2$ onto the $n=1$ states is always negligible, this is not the case in general as the states interact through non-adiabatic radial couplings. For example, the cross sections from the $n=3$ onto the $n=2$ states is not negligible, although the cross sections inside the $n=3$ manifold are not modified by the inclusion of the $n=2$ states. This means that the cross sections onto the $n=2$ states result from a decrease of the elastic cross section. 


The second observation is that there is a dependence of the cross section with $n$. This dependence was expected since the cross section scales classically as $n^4$ for Rydberg states. 

Finally, we also observed that, in a $n$ manifold, the cross section always increases with the orbital quantum number $l$ of the initial state.

\subsection{$n=2$ states}

\begin{figure}[h!]
\centering
\includegraphics[width=10cm]{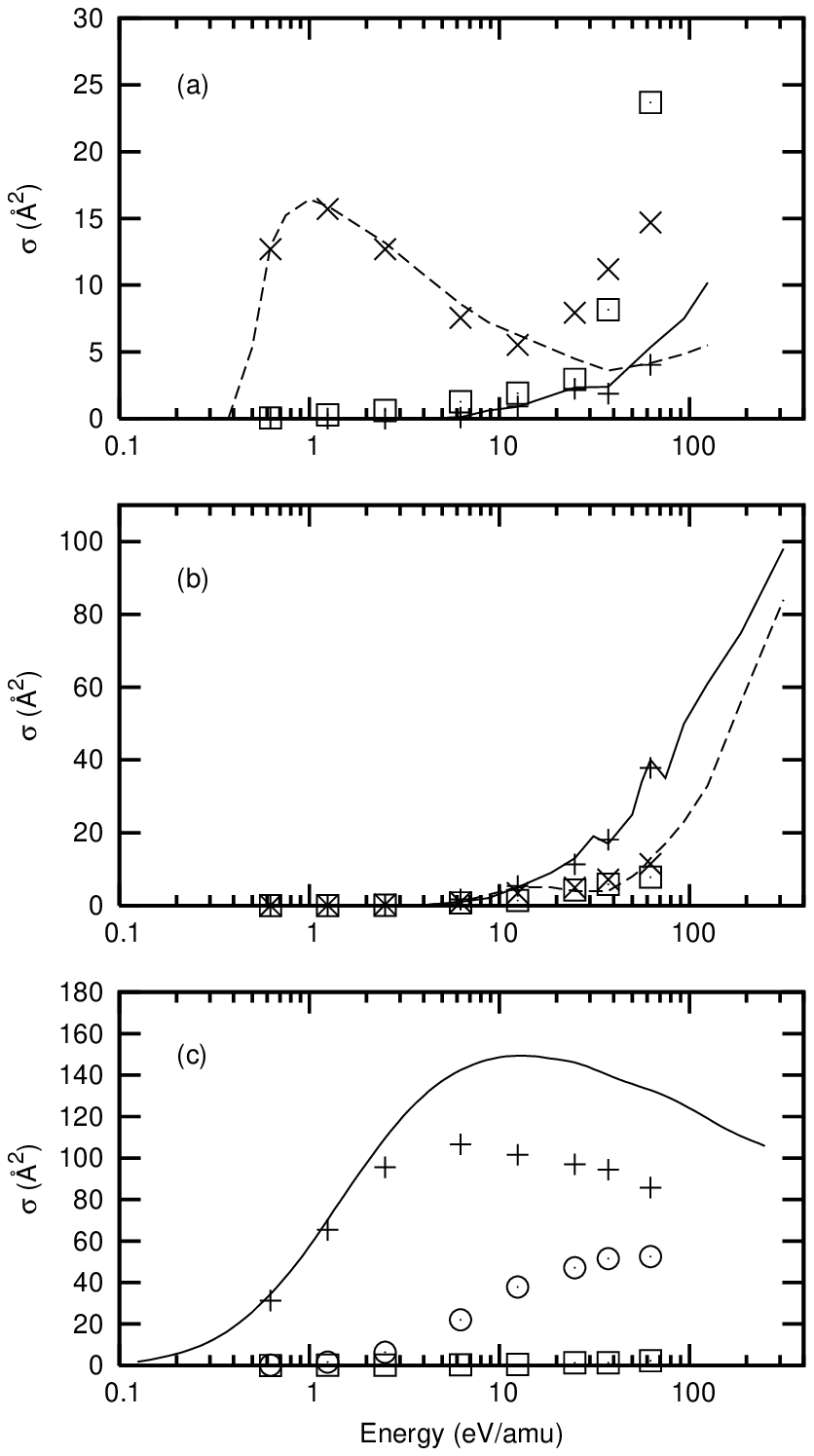}
\caption{Charge transfer cross sections between the $n=2$ states. \\
(a) With He$^+(1s)$ + H$(2s)$, $^1\Sigma^+$, as the initial state. Full line: charge transfer onto He$(1s2s\ ^1S)$ + H$^+$, $^1\Sigma^+$. $+$ signs: same, but with rotational couplings. Dashed line: charge transfer onto He$(1s2p\ ^1P^o)$ + H$^+$, $^1\Sigma^+$. $\times$ signs: same, but with rotational couplings. $\square$ signs: charge transfer onto He$(1s2p\ ^1P^o)$, $^1\Pi$. \\
(b) The same as (a), but with He$^+(1s)$ + H$(2p)$, $^1\Sigma^+$, as the initial state. \\
(c) With He$^+(1s)$ + H$(2s)$, $^1\Pi$, as the initial state. Full line: charge transfer onto He$(1s2p\ ^1P^o)$ + H$^+$, $^1\Pi$. $+$ signs: same, but with rotational couplings. $\square$ signs: charge transfer onto He$(1s2s\ ^1S)$ + H$^+$, $^1\Sigma^+$. $\bigcirc$ signs: charge transfer onto He$(1s2p\ ^1P^o)$ + H$^+$, $^1\Sigma^+$. }
\label{QCS_n=2}
\end{figure}

The cross sections with H$(2p)$ and H$(2s)$ in the $\Sigma$ symmetry as well as H($2p$) in the $\Pi$ symmetry are presented in figures \ref{QCS_n=2}.

The behavior of the cross section for the process H($2p$) + He$^+(1s) \rightarrow$ He($1s2p$) + H$^+$ is completely different in the $\Sigma$ (figure \ref{QCS_n=2}b) and $\Pi$ (figure \ref{QCS_n=2}c) symmetry: the total cross section from H($2p$) will be governed by the $\Pi$ states at low energy and by the $\Sigma$ states at high energy ($E\geq 100$ eV/amu).
Another difference is the behaviour of the cross section when the rotational couplings are included in the calculations: they have no effect for the transition with H($2p$) + He$^+(1s)$ in the $\Sigma$ symmetry as the initial state, but strongly modifiy the cross section for the corresponding $\Pi$ state. The cross section between the two $\Pi$ states is decreased while the cross section from $\Pi$ to $\Sigma$ states is increased so that the total cross section with H($2p$) + He$^+(1s)$ as the initial state is roughly the same as when the rotational couplings were neglected. 

The transition H($2s$) + He$^+(1s) \rightarrow$ He($1s2p$) + H$^+$ is also affected by the inclusion of the rotational couplings at energies $E\geq 10$ eV/amu (figure \ref{QCS_n=2}a). However, in this case the cross section between $\Sigma$ states and from $\Sigma$ to $\Pi$ states are both increased. This simply means that for this state, the elastic cross section is decreased by the inclusion of rotational couplings.

\subsection{$n=3$ states}

\begin{figure}[h!]
\centering
\includegraphics[width=10cm]{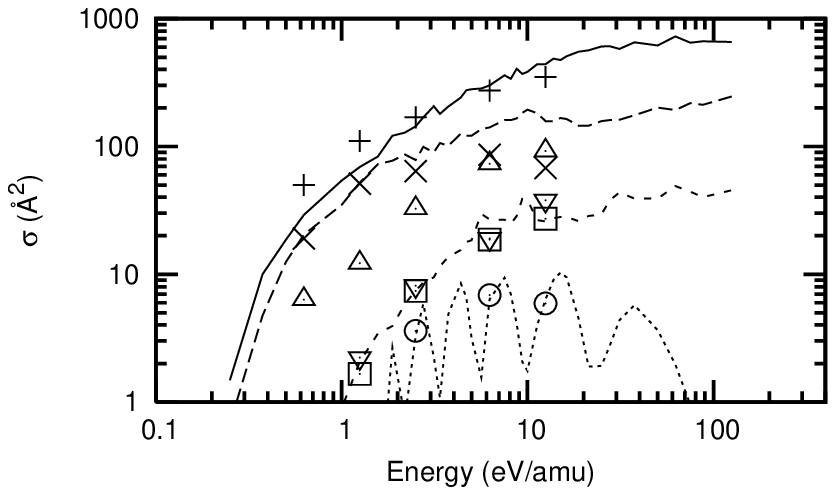}
\caption{Charge transfer cross sections with He$^+(1s)$ + H$(3d)$, $^1\Sigma^+$, as the initial state.
Full line: charge transfer onto He$(1s3s\ ^1S)$ + H$^+$, $^1\Sigma^+$. $+$ signs: same, but with rotational couplings.
Dashed line: charge transfer onto He$(1s3d\ ^1D)$ + H$^+$, $^1\Sigma^+$. $\times$ signs: same, but with rotational couplings.
Light dashed line: charge transfer onto He$(1s3p\ ^1P^o)$ + H$^+$, $^1\Sigma^+$. $\square$ signs: same, but with rotational couplings.
Dots: charge transfer onto He$(1s2p\ ^1P^o)$ + H$^+$, $^1\Sigma^+$. $\bigcirc$ signs: same, but with rotational couplings.
$\vartriangle$ signs: charge transfer onto He$(1s3d\ ^1D)$ + H$^+$, $^1\Pi$.
$\triangledown$ signs: charge transfer onto He$(1s3p\ ^1P^o)$ + H$^+$, $^1\Pi$.}
\label{QCSSigma8}
\end{figure}

\begin{figure}[h!]
\centering
\includegraphics[width=10cm]{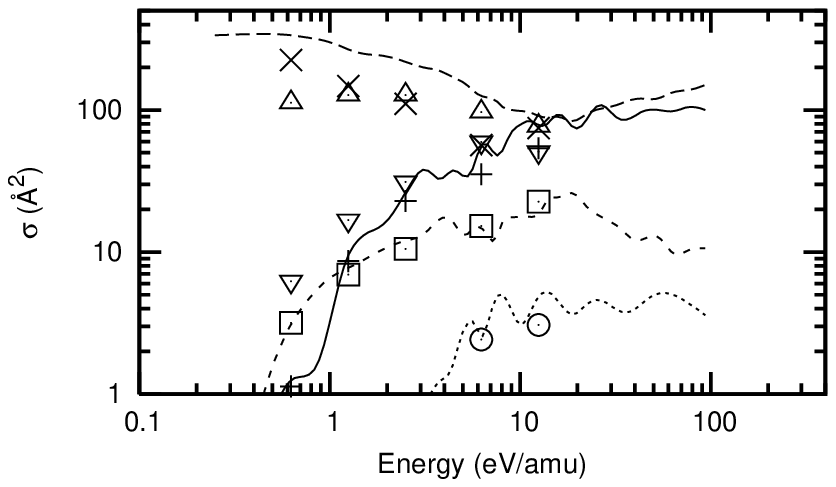}
\caption{Charge transfer cross sections with He$^+(1s)$ + H$(3p)$, $^1\Sigma^+$, as the initial state.
Full line: charge transfer onto He$(1s3s\ ^1S)$ + H$^+$, $^1\Sigma^+$. $+$ signs: same, but with rotational couplings.
Dashed line: charge transfer onto He$(1s3d\ ^1D)$ + H$^+$, $^1\Sigma^+$. $\times$ signs: same, but with rotational couplings.
Light dashed line: charge transfer onto He$(1s3p\ ^1P^o)$ + H$^+$, $^1\Sigma^+$. $\square$ signs: same, but with rotational couplings.
Dots: charge transfer onto He$(1s2p\ ^1P^o)$ + H$^+$, $^1\Sigma^+$. $\bigcirc$ signs: same, but with rotational couplings.
$\vartriangle$ signs: charge transfer onto He$(1s3d\ ^1D)$ + H$^+$, $^1\Pi$.
$\triangledown$ signs: charge transfer onto He$(1s3p\ ^1P^o)$ + H$^+$, $^1\Pi$.}
\label{QCSSigma10}
\end{figure}

\begin{figure}[h!]
\centering
\includegraphics[width=10cm]{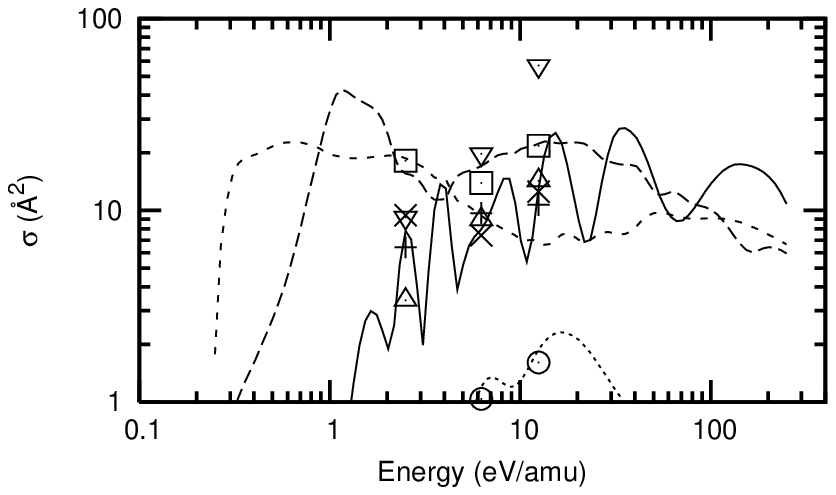}
\caption{Charge transfer cross sections with He$^+(1s)$ + H$(3s)$, $^1\Sigma^+$, as the initial state.
Full line: charge transfer onto He$(1s3s\ ^1S)$ + H$^+$, $^1\Sigma^+$. $+$ signs: same, but with rotational couplings.
Dashed line: charge transfer onto He$(1s3d\ ^1D)$ + H$^+$, $^1\Sigma^+$. $\times$ signs: same, but with rotational couplings.
Light dashed line: charge transfer onto He$(1s3p\ ^1P^o)$ + H$^+$, $^1\Sigma^+$. $\square$ signs: same, but with rotational couplings.
Dots: charge transfer onto He$(1s2p\ ^1P^o)$ + H$^+$, $^1\Sigma^+$. $\bigcirc$ signs: same, but with rotational couplings.
$\vartriangle$ signs: charge transfer onto He$(1s3d\ ^1D)$ + H$^+$, $^1\Pi$.
$\triangledown$ signs: charge transfer onto He$(1s3p\ ^1P^o)$ + H$^+$, $^1\Pi$.}
\label{QCSSigma12}
\end{figure}

There are 12 $n=3$ states. The cross section between the six $^1\Sigma^+$ states, presented in figures \ref{QCSSigma8}, \ref{QCSSigma10} and \ref{QCSSigma12}, have been calculated including the $n=2$ states since the cross section from the $n=3$ to the $n=2$ states is not negligible, as shown on these figures. There are also four $n=3$ $^1\Pi$ states and two $^1\Delta$ states. The charge transfer cross sections between these states are not presented here, but are available upon request to one of the authors, along with all the cross sections presented throughout this article.
From these figures, it is clear that there is a dependence of the cross section in the principal and orbital quantum numbers, $n$ and $l$, of the initial H($nl$) state: the charge transfer cross section is much larger for $n=3$ than for $n=2$, and also increases with the value of $l$.
The other difference between the $n=2$ and $n=3$ manifolds is the influence of rotational couplings. While in the $n=2$ states they were not influential at energies $E \leq 10$ eV/amu, this is not the case for the $n=3$ states where they play an important part even at energies below 1 eV/amu. We observe the intuitive fact that the  cross sections between $\Sigma$ states are smaller when the rotational interactions are taken into account, corresponding to the fact that a part of the cross section is transferred onto the $\Pi$ states.

We also observe that the cross sections onto the He($1s2p\ ^1P^o)$ + H$^+$ state in the $\Sigma$ symmetry is smaller when the initial state is higher in energy. The cross section to the other $n=2$ states, He($1s2s\ ^1S)$ + H$^+$ in the $\Sigma$ symmetry and He($1s2p\ ^1P^o)$ + H$^+$ in the $\Pi$ symmetry, are negligible and therefore not shown. Interestingly, the charge transfer cross sections from the $n=2$ states onto the $n=3$ states are all negligible.

\subsection{$n=4$ states}

\begin{figure}[h!]
\centering
\includegraphics[width=10cm]{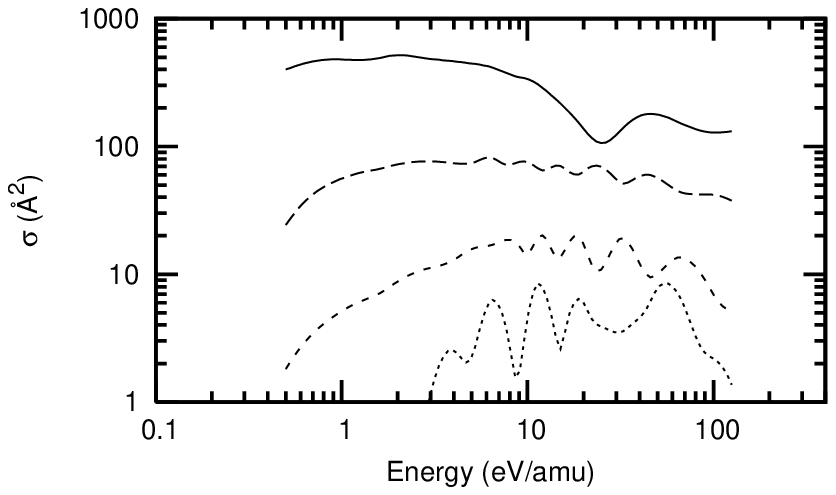}
\caption{Charge transfer cross sections with He$^+(1s)$ + H$(4p)$, $^1\Sigma^+$, as the initial state.
Full line: charge transfer onto He$(1s4s\ ^1S)$ + H$^+$, $^1\Sigma^+$. $+$ signs: same, but with rotational couplings.
Dashed line: charge transfer onto He$(1s3p\ ^1P^o)$ + H$^+$, $^1\Sigma^+$. 
Light dashed line: charge transfer onto He$(1s3d\ ^1D)$ + H$^+$, $^1\Sigma^+$. 
Dots: charge transfer onto He$(1s3s\ ^1S)$ + H$^+$, $^1\Sigma^+$.}
\label{QCSSigma14}
\end{figure}

We have included the first two $n=4$ singlet states in the $\Sigma$ symmetry. We could not consider more than the first two $n=4$ states in the diabatization since we were not able to calculate the radial non-adiabatic couplings for the higher-lying states. 
We calculated the cross section starting from the second $n=4$ state, H($4p$) + He$^+(1s)$. It confirms once again the dependence of the cross section in the quantum number $n$ for a given value of $l$. We also observe that the cross sections from this $n=4$ state onto the $n=3$ states are not negligible, as shown in figure \ref{QCSSigma14}. 

Moreover, it is observed (not shown) that despite the fact that the $n=3$ and $n=4$ states are close in energy and interact through radial non-adiabatic couplings, the cross sections in the $n=3$ manifold are not influenced by these two $n=4$ states, with the exception of the cross section with H($3s$) + He$^+(1s)$ as the initial state, which is slightly modified at energies higher than 10 eV/amu. In addition, the cross section from H($3s$) into the first $n=4$ state, He($1s4s$) + H$^+$, is negligible.
We thus reach the same conclusion as in the case of the $n=3$ states: the cross sections from the $n=3$ states onto the $n=2$ states are not negligible, although they do not influence the cross section within the $n=2$ manifold.

In conclusion, it would thus seems that the calculations of the charge transfer cross sections within a given $n$ manifold require to take into account only the states with $n^{\prime} \leq n$.

\newpage

\subsection{Total cross sections}\label{CS_tot}

The total cross sections starting from a given $nl$ state of H are obtained by summing all the contributions from within a $\Lambda$ manifold
\begin{equation}\label{QCS_total1}
\sigma (nl\Lambda) = \sum_{n^{\prime}=1}^{\infty} \sum_{l^{\prime}=0}^{n^{\prime}-1} \sigma(nl\Lambda \rightarrow n^{\prime}l^{\prime}\Lambda)
\end{equation}
and then by summing the contributions from all the $\Lambda$ \cite{Chibisov2002}:
\begin{equation}\label{QCS_total}
\sigma(nl) = \frac{1}{2l+1}\sum_{\Lambda=-l}^{l} \sigma(nl\Lambda)
\end{equation}
It should be noticed that as the states with $\Lambda\neq 0$ are doubly degenerate, they contribute twice in the sum in equation (\ref{QCS_total}). When the rotational couplings are taken into account, the only difference is an additional sum over $\Lambda^{\prime}$ in equation (\ref{QCS_total1}).

\begin{figure}[h!]
\centering
\includegraphics[width=12cm]{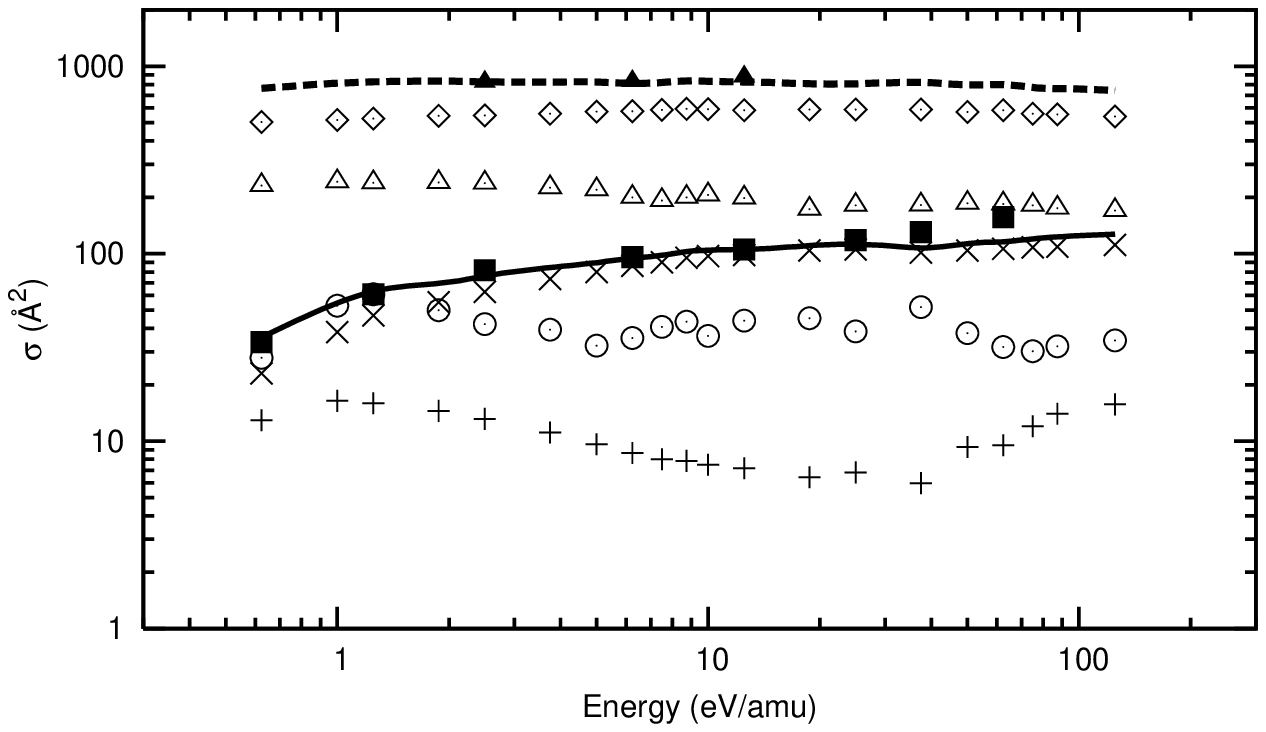}
\caption{Total charge transfer cross sections starting from H($nl$) + He$^+(1s)$. \\
$+$ signs: H($2s)$, $\times$ signs: H($2p$), Full line: H($n=2$), and $\blacksquare$ signs: H($n=2$) with rotational couplings. \\
$\bigcirc$ signs: H($3s$). $\vartriangle$ signs: H($3p$). $\lozenge$ signs: H($3d$). Dashed line: H($n=3$) and $\blacktriangle$ signs: H($n=2$) with rotational couplings.}
\label{QCS_tot}
\end{figure}

The total cross sections from the H($nl$) states, with $n=2,3$ are shown in figure \ref{QCS_tot} on a log-log scale. This figure clearly illustrates the dependence of the charge transfer cross section in $n$ and $l$. It also shows that while the inclusion of rotational couplings modifies the behavior of state-to-state cross sections at low energy, it modifies only slightly the total charge transfer cross section of a $n$ manifold. At energies higher than 10 eV/amu, the influence of rotational couplings starts to be important and the total cross section is increased. It would therefore be interesting to investigate the contributions of the rotational couplings at higher energies for $n=3$ states, but it is clear that our method is not adapted to such calculations.

\begin{figure}[h!]
\centering
\includegraphics[width=10cm]{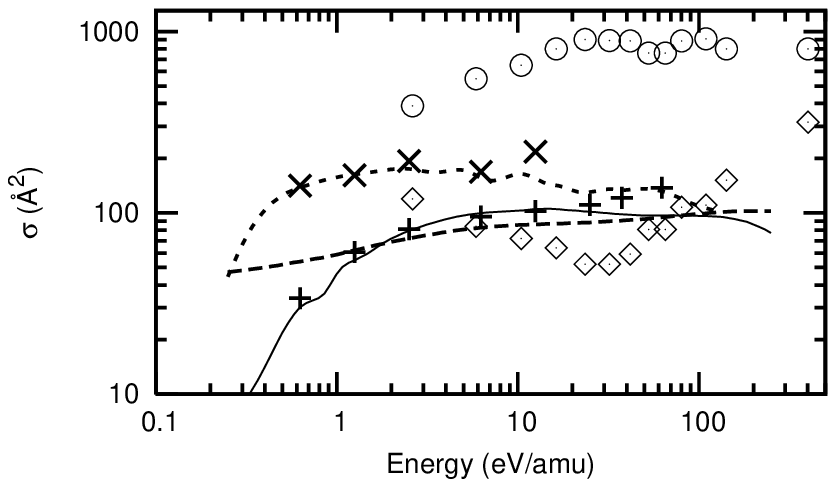}
\caption{Charge transfer cross sections with He($1s2p\ ^{1,3}P^o$) + H$^+$ and He($1s3p\ ^{1}P^o$) + H$^+$ as the final state and comparison with \cite{Chibisov2002}.
Full line: He($1s2p\ ^{1}P^o$) + H$^+$ as the final state. $+$ signs: same, but with rotational couplings. Dashed lines: He($1s2p\ ^{3}P^o$) + H$^+$ as the final state. Light dashed line: He($1s3p\ ^{1}P^o$) + H$^+$ as the final state. $\times$ signs: same, but with rotational couplings. Circles and diamonds: calculations of \cite{Chibisov2002} for the singlet and triplet states, respectively. The points were extracted graphically.}
\label{H2p_tot}
\end{figure}

In the same way, we can determine the total cross section with He$(1snl\ ^1L)$ + H$^+$ state as the final state. In helium-based plasma diagnostic, a correct estimation of the populations of the various He($1snl\ ^{1,3}L)$ levels, which are modified by charge transfer, is necessary. States such as He($1snp\ ^1P^o)$ decay radiatively to the ground state and these emission lines can be observed. In figure \ref{H2p_tot}, we have grouped together the charge transfer cross sections with He($1s2p\ ^{1,3}P^o)$ + H$^+$ and He($1s3p\ ^1P^o)$ + H$^+$ as the final states. Interestingly, the conclusions are similar when we consider the sum of all cross sections into a specific final state as when we considered a specific initial state: we again see a dependence in $n$, which is now the principal quantum number of the helium atom. We also see that the influence of the rotational couplings is weak at low energies but that the total charge transfer cross section is increased at energies $\geq10$ eV/amu.

These results can be compared with those of Chibisov {\it et al} \cite{Chibisov2002}. To describe the charge transfer process, these authors used a semi-classical method with an atomic basis where only the Stark couplings between the atomic states are taken into account and without rotational couplings.
The cross section with He($1s2p$) as the final state (\cite{Chibisov2002}, fig. 3) can be compared in the range of energy between $2.5$ and $200$ eV/amu (see figure \ref{H2p_tot}). For the singlet states, we see that not only is the behavior at small energies different, but also that the cross section is several times smaller in our calculations. In the triplet symmetry, the order of magnitude of the cross sections are roughly the same, but the general behavior is different.
The comparison can also be made for the $n=3$ states for $\Sigma$, $\Pi$ and $\Delta$. In \cite{Chibisov2001}, Chibisov {\it et al} present state-to-state cross section calculations, so that the comparison with our calculations is direct. The results are again qualitatively very different, showing the limitations of their method.

\section{Conclusions}

Using a quasi-molecular approach and a wave packet propagation method, we have computed the state-to-state cross sections for the charge transfer collisional process H($nl$) + He$^+(1s)$ $\rightarrow$ He($1sn^{\prime}l^{\prime}\ ^1L^{\prime})$ for all the $n,n^{\prime}=2,3$ singlet states (as well as the first two $n=4$ states in the $^1\Sigma^+$ symmetry) in the energy range between 0.25 and 150 eV/amu. We have also investigated the effect of the non-adiabatic rotational couplings on the charge transfer cross sections. All the cross sections are not presented in this article, but are available upon request to one of the authors.

We have found that our method is adapted when the rotational couplings are neglected, but is problematic at energies higher than 10 eV/amu when the rotational couplings are included due to very long computational times. 

We have found a strong dependence of the state-to-state charge transfer cross sections in the principal and orbital quantum numbers, $n$ and $l$, of the hydrogen atom. 
We observed that the rotational couplings have an influence on the cross sections even at low energies, but that their effect is increased with $n$: for $n=2$ states, we found that the effect of the couplings start to be important at energies higher than about 5 eV/amu, while for the $n=3$ states they modify the cross sections even at energies below 1 eV/amu. However, the total cross sections are not modified by the inclusion of rotational couplings at energies below 10 eV/amu. The effect of these couplings should therefore be investigated at intermediate energies by another method.

\section{Acknowledgments}
J. L. would like to thank the FRIA for financial support. This work was supported by the Fonds National de la Recherche Scientifique (IISN projects) and by the ÒAction de Recherche Concert\'eeÓ ATMOS de la Communaut\'e Fran\c caise de Belgique. The financial support of the FNRS in the University of Li\`ege Nic2 and Nic3 projects is gratefully acknowledged. We thank the support of the COST Action CM0702 CUSPFEL. KS would like to thank the University of Paris-Sud 11 for the postdoctoral fellowship N¡:2238.


\newpage
\pagebreak

\newpage
\pagebreak

\end{document}